\def\centereps#1#2#3{\vskip#2\relax\centerline{\hbox to#1{\special
  {eps:#3 x=#1, y=#2}\hfil}}}

\advance\voffset by -25mm
\advance\hoffset by -25mm

\documentclass{article}[11pt]
\usepackage[]{graphicx}
\newcounter{saveeqn}

\begin{document}
\null \vspace{30mm}
\setlength{\baselineskip}{20pt}

\noindent
\centerline{\LARGE { {\bf Is the phase of plane waves  an invariant?}}}
\vspace{10mm}

\centerline{Young-Sea Huang }
\centerline{Department of Physics, Soochow University, Shih-Lin, Taipei, Taiwan}
\centerline{yshuang@mail.scu.edu.tw}

\vspace{25mm}

\begin{abstract}
Based on the invariance of the phase of waves, plane waves was shown to propagate with {\it negative frequencies} in a medium which moves at superluminal speeds  opposite to the propagation direction of plane waves. The validity of the invariance of the phase of plane waves was then called into question. A radical change of the conventional concept of plane waves is recently proposed to solve the problem of negative frequency of waves. Here, we point out flaws in that proposal. Thus, the validity of the invariance of the phase of plane waves remains questionable. 
 
\end{abstract}

\vspace{4cm}



\vfill\eject

   The invariance of the phase of plane waves is an important concept in physics.  Is the phase of plane waves a frame-independent quantity? To almost all physicists, the answer is yes. Nevertheless, the invariance of the phase of plane waves has never been proved; it is only postulated, or argued, to be valid.\cite{jackson, rindler}. From the invariance of the phase of plane waves, the 4-vector $(\omega / c, {\bf k})$ of plane waves is shown to be Lorentz-covariant. Surprisingly, based on the Lorentz-covariance of the 4-vector $(\omega / c, {\bf k})$, plane waves are shown to propagate with {\it negative frequencies} in a medium which moves at superluminal speeds  opposite to the propagation direction of plane waves.\cite{huang} Furthermore, the Doppler effect can be derived without assuming the invariance of the phase of plane waves.\cite{Huang_Lu}  Thus, the validity of the invariance of the phase of plane waves is called into question.\cite{huang}

   Recently, a new concept of plane waves is proposed to resolve the problem of negative frequency of waves.\cite{Gjurchinovski}  It is argued that the apparent non-invariance of the phase of waves is due to an ignorance of the effect of relativistically-induced optical anisotropy because of the motion of the medium. By taking this effect into consideration, new forms of the "phase"  $ \Phi =  {\bf k} \cdot {\bf r} - {\bf k} \cdot {\bf u}\, t $  and of the frequency $ \omega = |{\bf k} \cdot {\bf u}| $ for plane waves are introduced to replace the conventional one $ \phi =  {\bf k} \cdot {\bf r} - \omega\, t $. The new velocity ${\bf u}$ of plane waves is only described as the velocity of the profile of plane waves; the  general definition of the new velocity ${\bf u}$ of plane waves is not precisely given. The proposal preserves the invariance of the "phase" of plane waves, but alters the conventional concept of plane waves. From the invariance of the new phase of plane waves and the Lorentz-covariant 4-vector $(ct, {\bf r})$, the relativistic transformations of ${\bf u}$, $ {\bf k}$,  and $\omega$ are obtained as \cite{Gjurchinovski} 
\begin{equation}\label{gj6}
 k_{x}  =  \gamma \, (k'_{x} + { V \over c^{2} }{\bf k}' \cdot {\bf u}'), \,\,\,\, k_{y}=k'_{y}, \,\,\,\, k_{z}=k'_{z},  
\end{equation}
\begin{equation}\label{gj7}
{ {\bf k} \cdot {\bf u}  \over c } = \gamma \, ( {k'_{x} V \over c } +  { {\bf k}' \cdot {\bf u}'  \over c } ),
\end{equation}
\begin{equation}\label{gj8}
u_{x}  = { u'_{x} +V \over 1+ {u'_{x} V  \over c^{2} } }, \,\,\,\, u_{y}  = { u'_{y}/ \gamma \over 1+ {u'_{x} V  \over c^{2} } }, \,\,\,\, u_{z}  = { u'_{z}/ \gamma \over 1+ {u'_{x} V  \over c^{2} } },
\end{equation}
\begin{equation}\label{gj9}
\omega' = | {\bf k}' \cdot {\bf u}'| , \,\,\,\, \omega = | {\bf k} \cdot {\bf u}| ,
\end{equation}
where ${\bf V}$ is the relative velocity along $x$-axis between frames, and $\gamma = 1/ \sqrt{ 1- (V/c)^{2} }$.   
For the particular case that the medium moves at superluminal speeds  opposite to the propagation direction of plane waves, the negative frequency problem is resolved in accordance with these relativistic transformations. Then, the "phase" of plane waves is claimed as a frame-independent quantity.

  Here, we will show that the proposal is false, and thus the question, {\it Is  the phase of plane waves an  invariant} ?, remains unsettled.  The induced anisotropy effect is a direct consequence of the Lorentz transformation. That is, the Lorentz transformation entirely takes into account this effect. No extra consideration of this effect is needed as  done usually.\cite{penfield,bladel,leonhardt,nandi}  Extra  consideration of this effect by the proposal not only induces conceptual problems on the characterization of plane waves, but also deduces false results. 

Conventionally, the wavefront normal of plane waves is clearly described by the wave vector ${\bf k}$, and the {\it phase velocity} is defined as ${\bf v} = (\omega / k)\, \hat{\bf k}$. Based on the definition of the phase velocity, one can certainly determine the phase velocity of waves propagating in the medium which is either at rest, or in motion.  
The new concept of plane waves by the proposal substantially alters the conventional concept of plane waves.  The description of the velocity ${\bf u}$ is clear only with the case that plane waves propagate in the medium-rest frame.   Based on the description of the velocity ${\bf u}$, could one determine the  velocity ${\bf u}$ of plane waves propagating in the medium which is in motion? 

\begin{figure}[h]
\begin{center}
\includegraphics[width=0.49\textwidth, clip=]{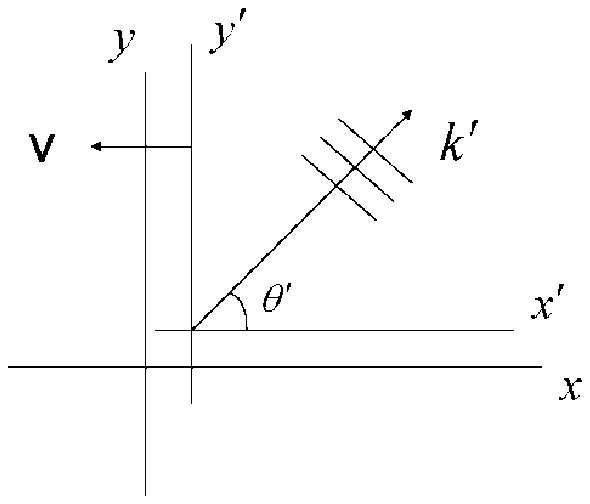}
\end{center}
\caption{\label{figa}
Monochromatic  waves  propagate with the wave vector ${\bf k}'$  relative to the frame  $S'$. The frame $S'$ is schematically represented by the $(x',y')$ coordinate system. The direction of the wave vector ${\bf k}'$ makes an angle $\theta '$ with the $x'-$axis of the frame  $S'$.  The  frame  $S'$ moves with a constant velocity $V$ along the negative $x$-axis direction, with respect to the frame $S$.}
\end{figure}

In the following, we explicitly point out fatal flaws in the new concept of plane waves. Referring to Fig.~\ref{figa}, consider monochromatic waves propagating in a medium at rest in the frame $S'$. 

This frame $S'$ moves with a constant velocity ${\bf V}$ along the negative $x$-axis direction with respect to the frame $S$. For simplicity, the wave vector ${\bf k}'$ of the waves is chosen to be parallel to the $(x',y')$ coordinate plane and it makes an angle $\theta '$ with the $x'-$axis. 
With respect to the medium-rest frame $S'$,  the wave vector ${\bf k}'$ and the velocity ${\bf u}'$ of the plane waves are in the same direction, and $u'= \omega' / k'$. From Eqs.~(\ref{gj6}) - (\ref{gj9}), we have
\begin{equation}\label{eqa1}
\cases{& $\!\!\!  k_{x}  =  \gamma (k'\, \cos \theta' - V\, \omega' / c^{2}   ),$\cr
       & $\!\!\! k_{y}  =  k'\, \sin \theta',$\cr
      & $\!\!\! {\bf k} \cdot {\bf u} =   \gamma \, \omega'  (1 - V\, \cos \theta' / u'  ),$\cr  }
\end{equation}
and
\begin{equation}\label{eqa2}
\cases{& $\!\!\! u_{x}  =  {u' \cos \theta' -V  \over  1- u' \cos \theta' \, V / c^{2}  } ,$\cr
       & $\!\!\! u_{y}  =  {u' \sin \theta'  \over  \gamma \, ( 1- u' \cos \theta' \, V / c^{2} ) } .$\cr
      }
\end{equation}
Then, from Eqs.~(\ref{eqa1}) and (\ref{eqa2}), we obtain
\begin{equation}\label{eqa3}
{ k_{y} \over k_{x} } = {  \sin \theta'  \over \gamma \, (  \cos \theta' - V \, u'/  c^{2} ) },
\end{equation}
and
\begin{equation}\label{eqa4}
{ u_{y} \over u_{x} } = {  \sin \theta'  \over \gamma \, ( \cos \theta' - V /  u' ) }.
\end{equation}
 According to Eqs.~(\ref{eqa3}) and (\ref{eqa4}), with respect to the frame $S$, the wave vector ${\bf k}$ and the velocity ${\bf u}$ of the  plane waves are in general not along the same direction, except in the case that light waves propagate in vacuum, since $u' =c$. Without first knowing the wave vector ${\bf k'}$ and the velocity ${\bf u'}$ for the waves in the medium-rest frame $S'$, we can determine the phase velocity  ${\bf v}$ of the plane waves with respect to the frame $S$, relative to which  the medium is in motion. Contrarily, the given description of the velocity ${\bf u}$ by the proposal is incapable of enabling us to determine the velocity ${\bf u}$ of the plane waves with respect to the frame $S$, without first knowing the wave vector ${\bf k'}$ and the velocity ${\bf u'}$ for those waves in the medium-rest frame $S'$. The given description of the velocity ${\bf u}$ of plane  waves is  applicable only for the case that plane waves  propagate in the medium-rest frame.

 Now, let us examine the transverse Doppler effect in accordance with the new concept of plane waves. To find the frequency shift of  plane waves propagating transversely with respect to the frame $S$, which criterion do we apply to judge whether or not  plane waves propagate transversely: plane waves propagate along the direction of  ${\bf u}$, or along the direction of ${\bf k}$? According to the new concept of plane waves, plane waves propagate with the velocity ${\bf u}$, not with the phase velocity ${\bf v} = (\omega / k)\, \hat{\bf k}$. In this sense, we let $u_{x}=0$. Then, from Eqs.~(\ref{eqa1}) and (\ref{eqa2}), the frequency of the transverse plane waves with respect to the frame $S$ is obtained as $\omega = |{\bf k} \cdot {\bf u} | = \gamma\, | 1 -  ( V/ u' )^{2} | \,\omega '$, where $\omega '$ is the frequency of those plane waves as observed in the medium-rest frame $S'$. Instead, suppose that we  choose the other way, that  plane waves propagate along the direction of the wave vector ${\bf k}$. Then, let $k_{x}=0$. Similarly, from Eqs.~(\ref{eqa1}) and (\ref{eqa2}) the  frequency of the transverse plane waves with respect to the frame $S$ is $\omega = |{\bf k} \cdot {\bf u}| = \omega' / \gamma$. This result gives the same transverse Doppler shift as that predicted by special relativity, and has been confirmed experimentally. Thus, that plane waves propagate with the velocity ${\bf u}$ as suggested by the new concept of plane waves of the proposal is not valid.

 The new concept of plane waves of the proposal not only has conceptual problems on the characterization of plane waves, but also deduces false results on the transverse Doppler effect. Hence, the problem of negative frequency of waves is not resolved by the proposal. The validity of the invariance of the phase of plane waves remains questionable.

\end{document}